\documentclass [12pt,preprint]{aastex}
\usepackage{epsfig}
\begin{document}
\voffset-1cm
\newcommand{\gsim}{\hbox{\rlap{$^>$}$_\sim$}}
\newcommand{\lsim}{\hbox{\rlap{$^<$}$_\sim$}}

\title{On the origin of the correlations between \\
Gamma-Ray Burst observables}

\author{Shlomo Dado\altaffilmark{1}, Arnon Dar\altaffilmark{1}
and A. De 
R\'ujula\altaffilmark{2}}

\altaffiltext{1}{dado@phep3.technion.ac.il, arnon@physics.technion.ac.il,
dar@cern.ch.\\
Physics Department and Space Research Institute, Technion, Haifa 32000,
Israel}
\altaffiltext{2}{alvaro.derujula@cern.ch; Theory Unit, CERN,
1211 Geneva 23, Switzerland \\ 
Physics Department, Boston University, USA}

\begin{abstract}
Several pairs of observable properties of Gamma Ray Bursts (GRBs) 
are known to be correlated. 
Many such correlations are straightforward predictions of the 
`cannonball' model of GRBs. 
We extend our previous discussions of the subject to a wealth of new data, 
and to correlations between `lag-time', `variability' and `minimum rise-time',
with other observables. Schaefer's recent systematic analysis of the
observations of many GRBs of known red-shift gives us a good and
updated data-basis for our study.
\end{abstract}

\keywords{Gamma Ray Burst}

\section{Introduction}

Quite a few independent observable quantities can be measured
in a long-duration gamma-ray burst (GRB). These include its
 spherical equivalent energy,
 its peak isotropic luminosity,
the `peak energy' of its spectrum
and the red-shift of its host galaxy.
In order of increasing effort on the data analysis, one can also define 
and determine the number of pulses in the GRB's light-curve and
their widths, rise-times, and `lag-times'.
Finally, with considerable toil and embarrassment of choices,
one can define and measure `variability' 
(Fenimore  \& Ramirez-Ruiz, 2000; Plaga, 2001;
Reichart et al.~2001; 
Guidorzi et al.~2006; Schaefer 2006).
Pairs of the above quantities are known to be correlated
as approximate power laws, sometimes fairly
tightly and over spans of several orders of magnitude. 
A model of GRBs ought to be able to predict these power laws
and to pin-point the choices of red-shift corrections that should 
make them tightest.

In the `cannonball' (CB) model of GRBs the correlations
between the cited observables, as we shall discuss, are
predictable. They are based on very simple physics:
the production of $\gamma$ rays by inverse Compton scattering
of much softer photons by a relativistically-moving, 
quasi-point-like object (Shaviv \& Dar 1995, Dar \& De R\'ujula 2004).
The correlations very satisfactorily test 
these individual theoretical ingredients, or combinations thereof.
A reader desiring to start by evaluating these claims may 
choose to read Section \ref{basis} first.

The data, particularly on
GRBs of known red-shift, has become much more extensive
in the time elapsed since the CB-model correlations were
predicted (Dar \& De R\'ujula 2001) and several of them
were tested (Dar \& De R\'ujula 2004). It is time
to restudy the subject, which we do for many correlations,
relying mainly on the data analysis by Schaefer (2006).

\section{The CB model} 

In the CB model (Dar \& De R\'ujula~2000, 2004; Dado et al.~2002, 2003), 
{\it long-duration} GRBs 
and their AGs are produced by bipolar jets of CBs, ejected in 
core-collapse SN explosions (Dar \& Plaga~1999). An 
accretion disk  is hypothesized to be produced around the newly 
formed compact object, either by stellar material originally close to the 
surface of the imploding core and left behind by the explosion-generating 
outgoing shock, or by more distant stellar matter falling back after its 
passage (De R\'ujula~1987). As observed in microquasars, each 
time part of the disk falls abruptly onto the compact object, a 
pair of CBs made of {\it ordinary plasma} are emitted with high 
bulk-motion Lorentz factors, $\gamma$, in opposite directions along the 
rotation axis, wherefrom matter has already fallen onto the compact 
object, due to lack of rotational support. 
The $\gamma$-rays of a single 
pulse in a GRB are produced as a CB coasts through the SN {\it glory} 
--the SN light scattered away from the radial direction
by the SN and pre-SN ejecta. The electrons enclosed in the 
CB Compton up-scatter glory's photons to GRB energies. 

Each pulse of a GRB 
corresponds to one CB. The emission times of the individual CBs
reflect the chaotic accretion process and are not predictable.
At the moment,
neither are the characteristic baryon number and Lorentz factor
of CBs, which can be inferred from the analysis of GRB afterglows
(Dado et al.~2002, 2003a). Given this information,  
two other `priors' (the typical early
luminosity of a supernova and the typical density
distribution of the parent star's
wind-fed circumburst material), and a single extra hypothesis
(that the wind's column density in the `polar' directions 
is significantly smaller than average) all observed properties of 
the GRB pulses can be derived (Dar \& De R\'ujula~2004). 
All that is required are explicit simple calculations involving
Compton scattering. 

Strictly speaking, our results refer to single pulses in a GRB, whose 
properties reflect those of {\it one} CB. The statistics on single-pulse 
GRBs of known redshift are too meager to be significant.  Thus, we
apply our results to entire GRBs, irrespective of their number of pulses,
which ranges from 1 to $\sim\!12$. This implies averaging the properties
of a GRB over its distinct pulses,
and is no doubt a source of dispersion in the
correlations we study.

%The rapid expansion of the CBs stops shortly after ejection 
%(Dado et al.~2002, Dar \& De R\'ujula 2006) by their 
%interaction with the inter-stellar medium (ISM). During this  initial
%rapid expansion and cooling phase, their AG is  
%dominated by thermal bremstrahlung and line emission. Later, 
%their AG becomes dominated by synchrotron radiation from 
%swept-in ISM electrons spiraling in the CBs' inner magnetic fields
%(Dado et al.~2002, 2006) and from ISM electrons scattered to higher 
%energies by the moving CBs (Dado \& Dar 2005).

\section{The basis of the correlations, and a summary of results}
\label{basis}

Cannonballs 
are highly relativistic, their typical Lorentz factors are 
$\gamma\!=\!{\cal{O}}(10^3)$. They 
are quasi-point-like: the angle a CB subtends from its point
of emission is  comparable or
smaller than the characteristic opening angle, 
$1/\gamma$, of its relativistically beamed radiation.
Let  the typical viewing angle of an 
observer of a CB, relative to its direction of motion,
 be $\theta\!=\!{\cal{O}}$(1 mrad), and let 
$\delta\!=\!{\cal{O}}(10^3)$ be the corresponding Doppler factor:
\begin{equation}
\delta \equiv {1\over\gamma\,(1-\beta\, cos\theta)}
                       \simeq  {2\, \gamma
                       \over 1+\gamma^2\, \theta^2}\; ,
\label{delta}
\end{equation} 
where the approximation is excellent 
for  $\theta\ll 1$ and  $\gamma \gg 1$. 

To correlate two GRB observables, all one needs to know is
their functional dependence on $\delta$ and $\gamma$. The reason
is that, in Eq.~(\ref{delta}),
the $\theta$ dependence of $\delta(\gamma,\theta)$  is so pronounced, that 
it may be expected to be the
largest source of the case-to-case spread in the measured quantities
(for GRBs of known redshift $z$, the correlations are sharpened
by use of the explicit $z$-dependences). In the CB model,
the $(\gamma,\delta,z)$ dependences of the  
 spherical equivalent energy of a GRB, $E_\gamma^{\rm iso}$; 
 its peak isotropic luminosity $L_p^{\rm iso}$;
its peak energy, $E_p$ (Dar \& De R\'ujula 2001);
and its pulse rise-time $t_{\rm rise}$ (Dar \& De R\'ujula 2004), are:
\begin{equation}
E_\gamma^{\rm iso}\propto\delta^3,\;\;\;
(1+z)^2\,L_p^{\rm iso}\propto \delta^4,\;\;\;
(1+z)\,E_p\propto \gamma\,\delta,\;\;\;
t_{\rm rise}/(1+z)\propto 1/(\gamma\,\delta),
\label{brief}
\end{equation}
The first two of these results are simple consequences of relativity
and the quasi-point-like character of the CB-model's sources
(they would be different for an assumed GRB-generating
jet with an opening angle much greater than $1/\theta$). The expression for
$E_p$ reflects the inverse Compton scattering by the CB's electrons
(comoving with it with a Lorentz factor $\gamma$) of the glory's photons,
that are approximately isotropic in the supernova rest system, and
are Doppler-shifted by the CB's motion by a factor $\delta$ 
(the result would be different, for instance, for synchrotron radiation
from the GRB's source, or self-Compton scattering of photons
comoving with it). The expression for the pulses's rise-time has the
same physical basis as that for $E_p$, but we shall see in more
detail in Section \ref{variability}, Eq.~(\ref{twtrans22}), that it also reflects
the production of $\gamma$-rays in an illuminated, previously 
wind-fed medium\footnote{The coefficients of
proportionality in Eqs.~(\ref{brief}) have explicit dependences on 
the number of CBs in a GRB,  their
initial expansion velocity and baryon number.
With typical values fixed by the analysis of GRB afterglows,
the predictions agree with the
observations (Dar \& De R\'ujula 2004, Dado et al.~2006).}.

In Sections \ref{variability} and \ref{lag}
 we derive the CB-model's expectation for the
the variability of a GRB, $V$, and the prediction for the
`lag-time' of  its pulses, $t_{\rm lag}$, to wit:
\begin{equation}
V\propto \gamma\,\delta/(1+z),\;\;\;
t_{\rm lag}\propto (1+z)^2/(\delta^2\,\gamma^2) .
\label{tlagV}
\end{equation}
The physics of the first of these relations is essentially the same as
that of a pulse's rise-time. The behaviour of $t_{\rm lag}$ reflects
the CB-model's  specific prediction for how, as a pulse evolves in time, 
the photon's energy spectrum softens, due to the increasingly non-isotropic
character of the glory's photons which the CBs encounter as they travel,
and to the softening of the energy spectrum
of the CB's electron population 
(Dar \& De R\'ujula 2004).

Given the 6 relations in Eqs.~(\ref{brief},\ref{tlagV}), it is straightforward
to derive the 15 ensuing two-observable correlations, of which subgroups
of 5 are non-redundant (if $A$ is correlated to $B$ and $B$ to $C$...).
We consider first one of these subgroups: the 5 correlations 
most often phenomenologically 
discussed to date. 
All these correlations are derived in the same manner. Consequently,
we proceed by way of example and outline only the derivation of the 
$[E_p\,,E_\gamma^{\rm iso}]$ correlation
(Dar \& De R\'ujula 2001,  Dado et al.~2006). The full derivation is
in the Appendix.

\begin{figure}[]
\centering
\vbox{
\hbox{
 \epsfig{file=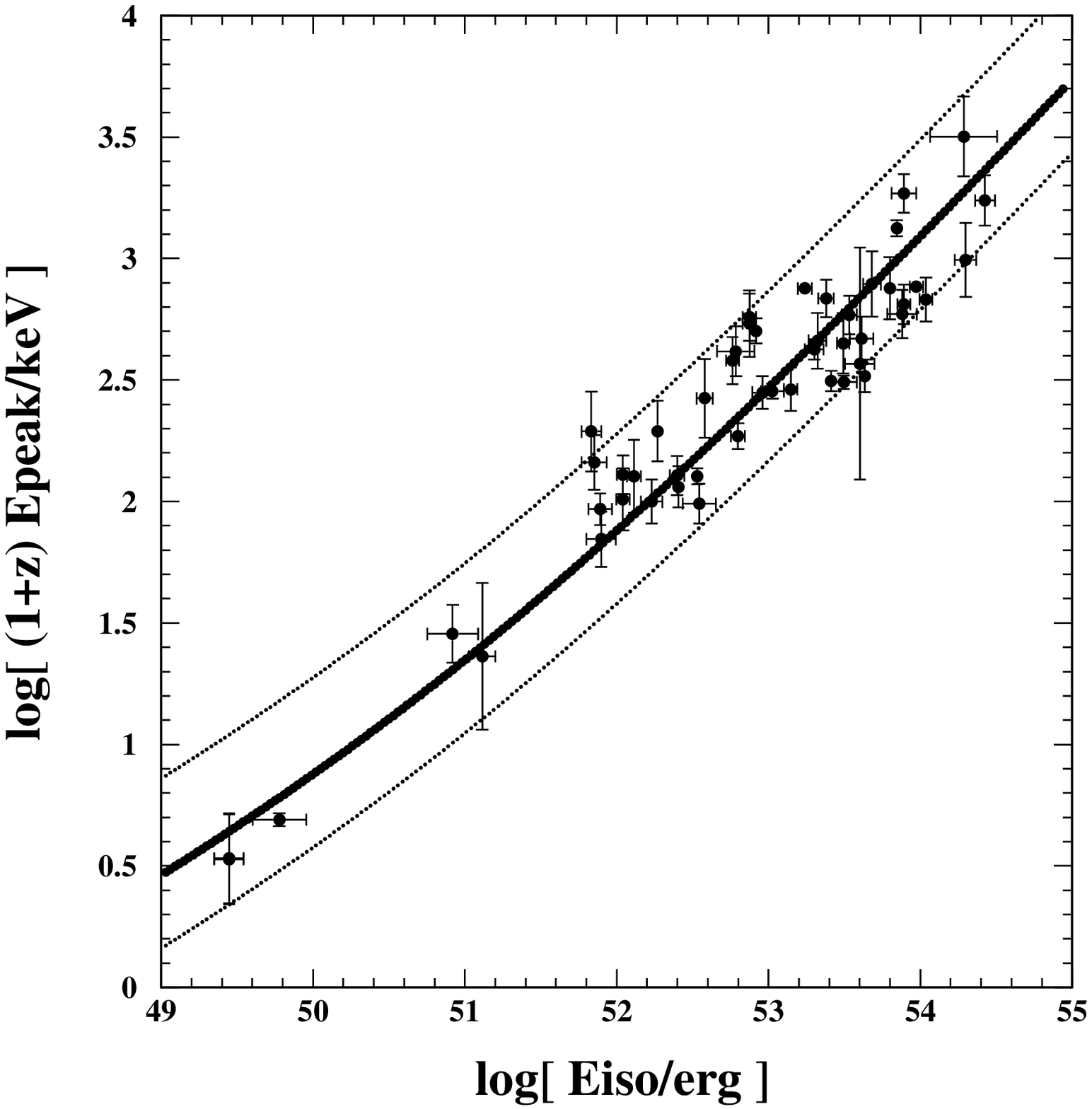,width=8cm}
 \epsfig{file=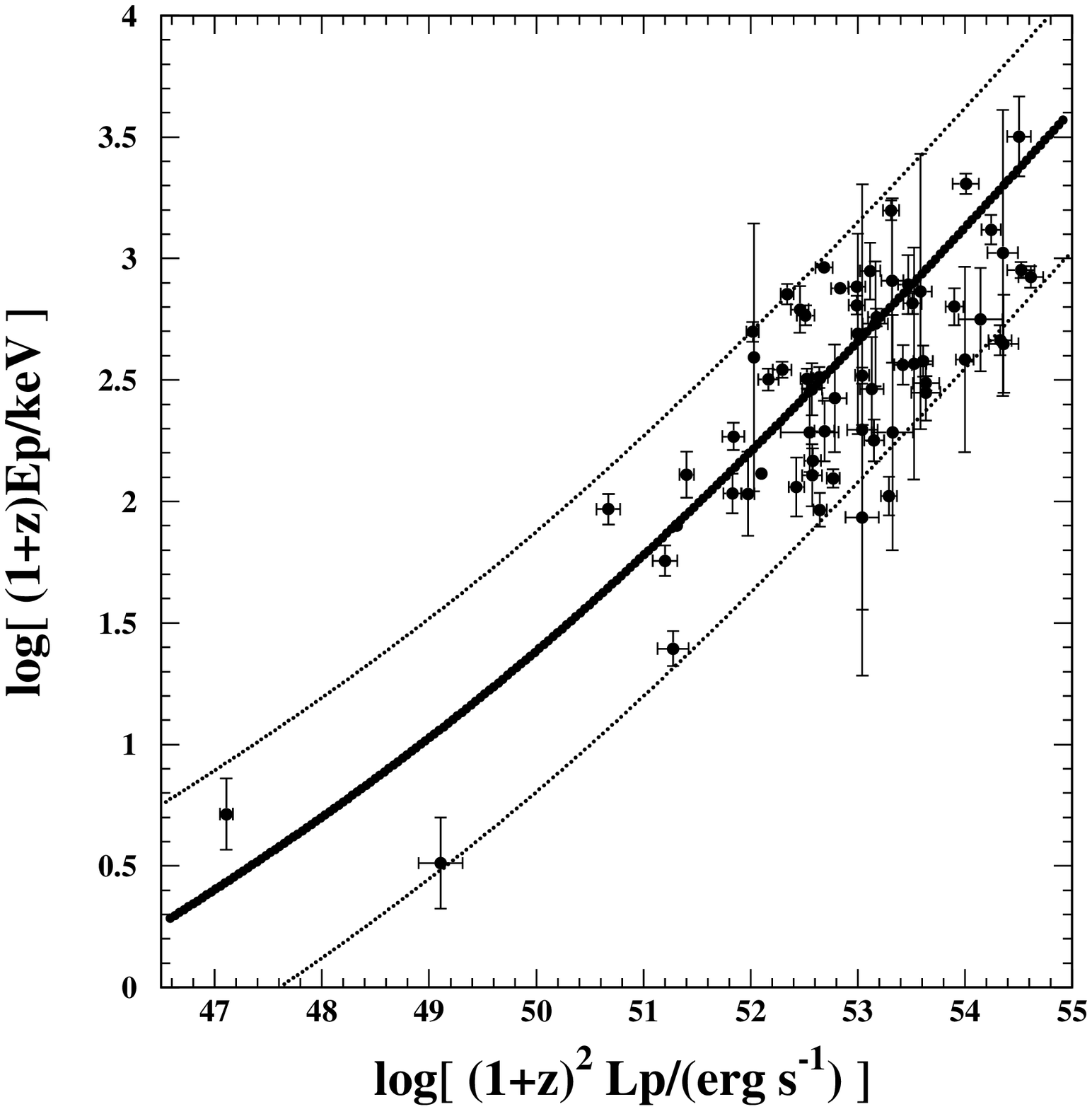,width=8cm}
}}
\vbox{
\hbox{
\epsfig{file=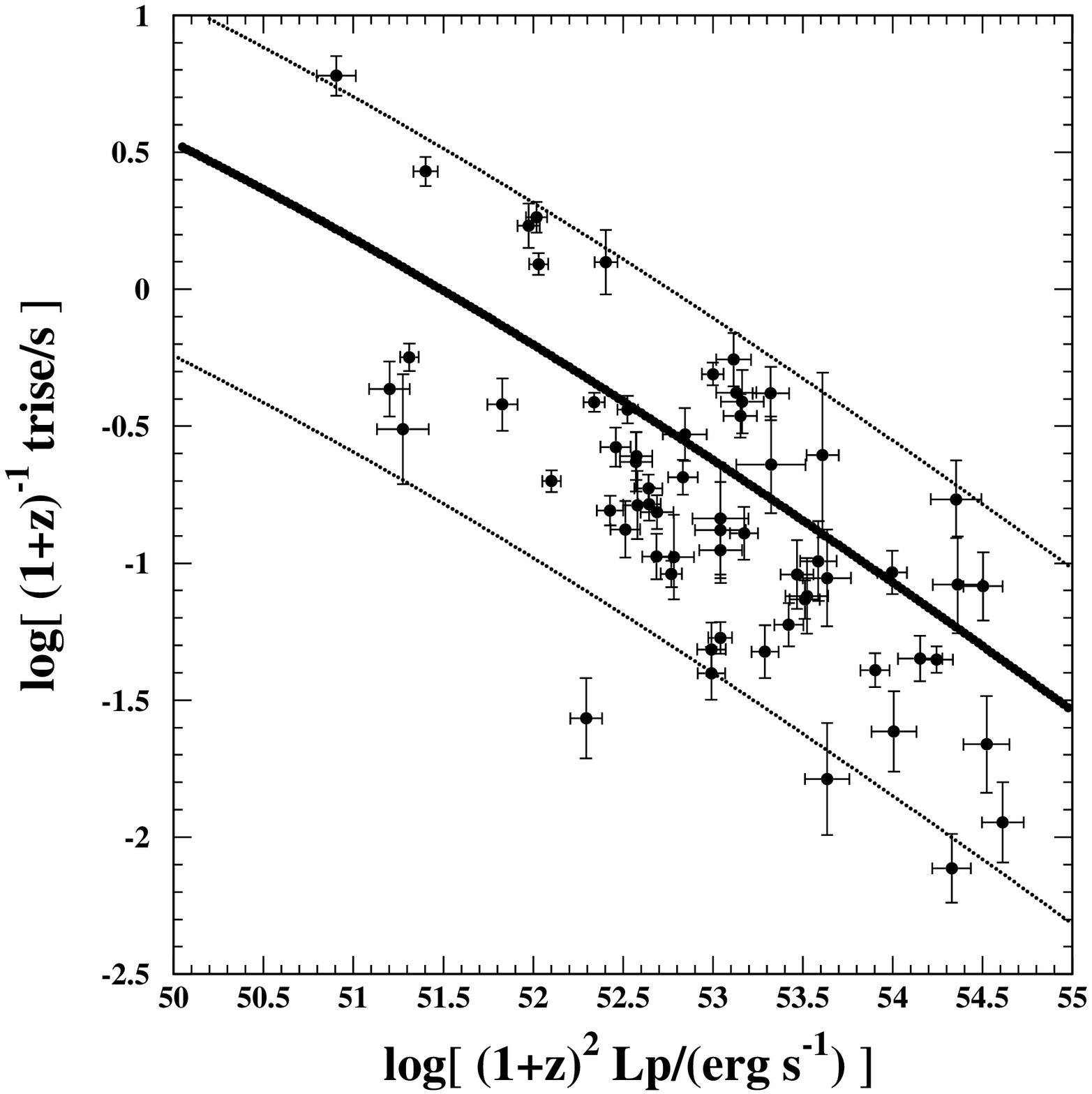,width=8cm}
\epsfig{file=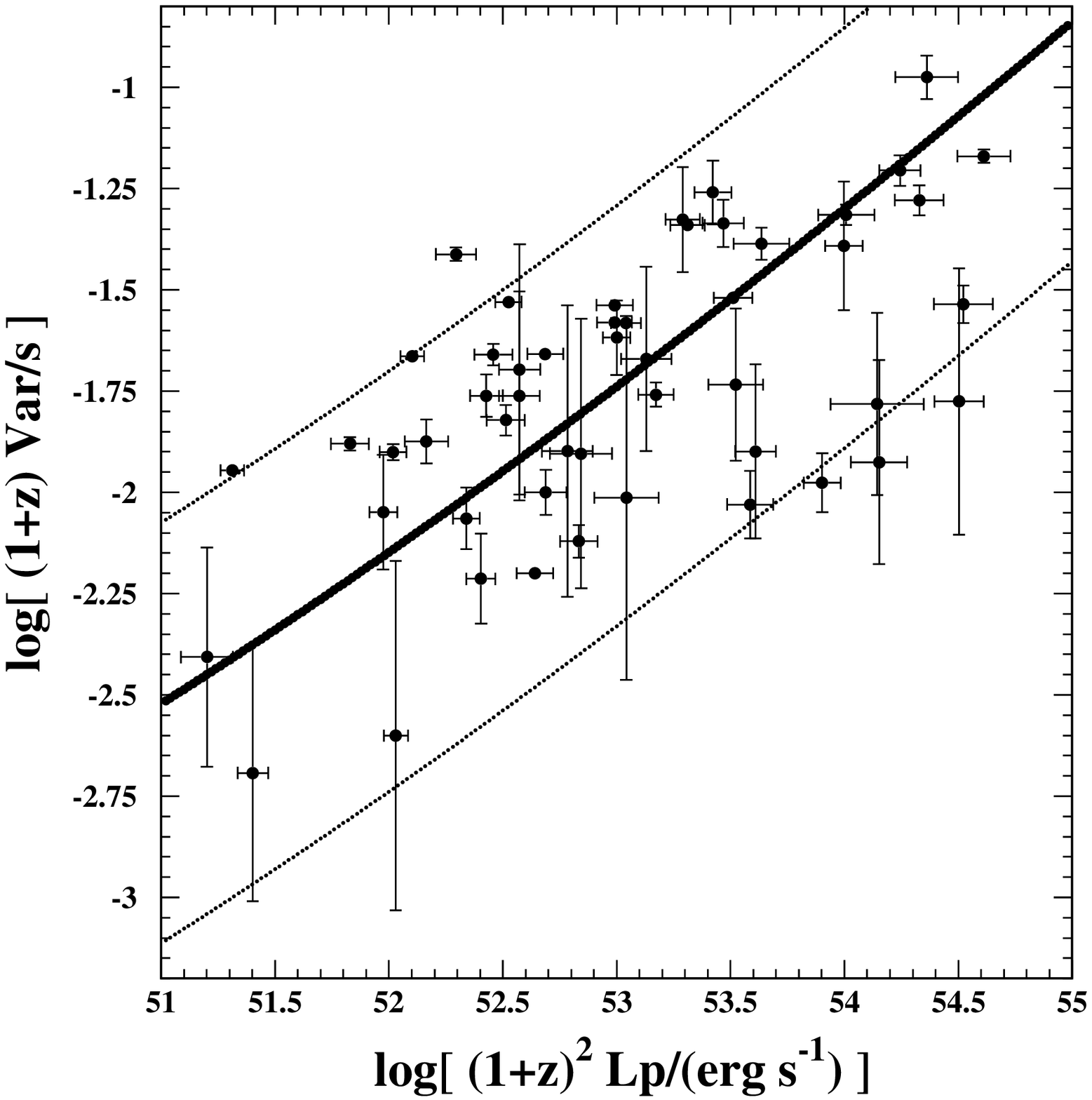,width=8cm}
}}
\vspace*{8pt}
\caption{
Left to right and top to bottom (a to d):
a) The $[E_p,E_\gamma^{\rm iso}]$ correlation of Eq.~(\ref{epi}).
b) The $[E_p,L_p^{\rm iso}]$ correlation of Eq.~(\ref{lpep}).
c) The $[t_{\rm min}^{\rm rise},L_p^{\rm iso}]$ correlation of
Eq.~(\ref{tmineq}).
d) The $[V,L_p^{\rm iso}]$ correlation of Eq.~(\ref{Veq}).
The dotted `variance lines' are to guide the eye, they are not
always symmetric about the best fit.
}
 \label{f1}
\end{figure}

An $[E_p\,,E_\gamma^{\rm iso}]$ 
correlation was predicted [and tested]
in Dar \& De R\'ujula (2001, [2004]).
According to Eqs.~(\ref{brief}), 
$(1\!+\! z)\,E_p\!\propto\!\gamma\delta$
and $E_\gamma^{\rm iso}\propto\delta^3$. If most of the variability
is attributed to the very fast-varying $\theta$-dependence of $\delta$ in
Eq.~(\ref{delta}),
$(1\!+\! z) E_p\!\propto\![E_\gamma^{\rm iso}]^{1/3}$. 
(Dar \& De R\'ujula 2001).
This original expectation can be refined
by exploiting another prediction (Dado et al.~2006).
A typical observer's
angle is $\theta\!\sim\!1/\gamma$. A relatively large $E_p$
implies a relatively large $\delta$, and a relatively small viewing angle,
$\theta<1/\gamma$. For $\theta^2 \ll 1/\gamma^2$, 
$\delta\propto\gamma$, implying that $(1\!+\! z)E_p\propto 
[E_\gamma^{\rm iso}]^{2/3}$ {\it for the largest observed values} of 
$E_\gamma^{\rm iso}$. On the other hand, for $\theta^2 \gg 1/\gamma^2$,
the Dar \& De R\'ujula (2001) correlation is unchanged: 
it should be increasingly accurate 
{\it for smaller values} of $E_\gamma^{\rm iso}$.
We may interpolate between these extremes by positing:
\begin{equation}
 (1\!+\! z)\,E_p=E_p^0\,\left\{
 [E_\gamma^{\rm iso}/E_0^{\rm iso}]^{1/3}
 +[E_\gamma^{\rm iso}/E_0^{\rm iso}]^{2/3}\right\} \; ,
\label{epi}
\end{equation}
an expression with two parameters ($E_p^0$, $E_0^{\rm iso}$);
like the correlation $E_p=a\,[E_\gamma^{\rm iso}]^b$
(see, e.g., Amati~2006a,b,c), whose power behaviour 
is arbitrary. A fit to Eq.~(\ref{epi}) is shown in Fig.~\ref{f1}a.
The variances around the
mean trends of all the correlations  we study
have roughly log-normal distributions. Thus our fits are to
the logarithms of the observed quantities.

The number fluence of a GRB is proportional
to $\delta^2$, its energy fluence to $\delta^3$. The individual-photon
energies are $\propto\delta\,\gamma$. All these facts imply that one
expects most observed events to correspond to small $\theta$ (though
obviously not to $\theta\simeq 0$, a set of null solid angle). Small
$\theta$ means large $E_p$ and $E_\gamma^{\rm iso}$, approximately 
related by $(1\!+\! z)E_p\propto [E_\gamma^{\rm iso}]^{2/3}$.
The expectation is supported by the data  in Fig.~\ref{f1}a.  
These comments
on the $[E_p\,,E_\gamma^{\rm iso}]$ 
correlation and the way it is derived are extensible to
the other correlations we shall discuss, e.g., to the
$[E_p\,,L_p^{\rm iso}]$ correlation predicted and tested
in Dar \& De R\'ujula (2001, 2004). Given Eq.~(\ref{brief})
and in analogy with Eq.~(\ref{epi}),
we expect (Dado et al.~2006):
\begin{equation} 
(1+z)\,E_p \simeq E_p^0\, 
\left\{[(1+z)^2\,L_p^{\rm iso}/L_p^0]^{1/4} + 
[(1+z)^2\,L_p^{\rm iso}/L_p^0]^{1/2}\right\}  .
\label{lpep}  
\end{equation}
This correlation is akin to the one proposed by  Yonetoku et al.~(2004),
but its power behaviour is not arbitrary.

To derive the other correlations we shall confront with data, we must 
recall the CB-model
expectations for pulse rise-times and event variabilities,
and derive the one for the lag-time. But Eqs.~(\ref{brief},\ref{tlagV})
allow us to anticipate the results:
\begin{equation}
{t_{\rm min}^{\rm rise}\, (1+z)^{-1}}=t_0  \left\{
\left[(1+z)^2\,L_p^{\rm iso}/L_p^0 \right]^{1\over 4}
+\left[(1+z)^2\,L_p^{\rm iso}/L_p^0\right]^{1\over 2} \right\}^{-1}
\label{tmineq}
\end{equation}
\begin{equation}
{(1+z)\,V}=V_0\;\left\{
\left[(1+z)^2\,L_p^{\rm iso}/L_p^0\right]^{1\over 4}
+\left[(1+z)^2\,L_p^{\rm iso}/L_p^0\right]^{1\over 2}\right\}
\label{Veq}
\end{equation}
\begin{equation}
{t_{\rm lag}\, (1+z)^{-2}}=t_0  \left\{
\left[(1+z)^2\,L_p^{\rm iso}/L_p^0 \right]^{1\over 2}
+\left[(1+z)^2\,L_p^{\rm iso}/L_p^0\right]^{1} \right\}^{-1}
\label{tlageq}
\end{equation}

\begin{figure}[]
\centering
\vbox{
\hbox{
 \epsfig{file=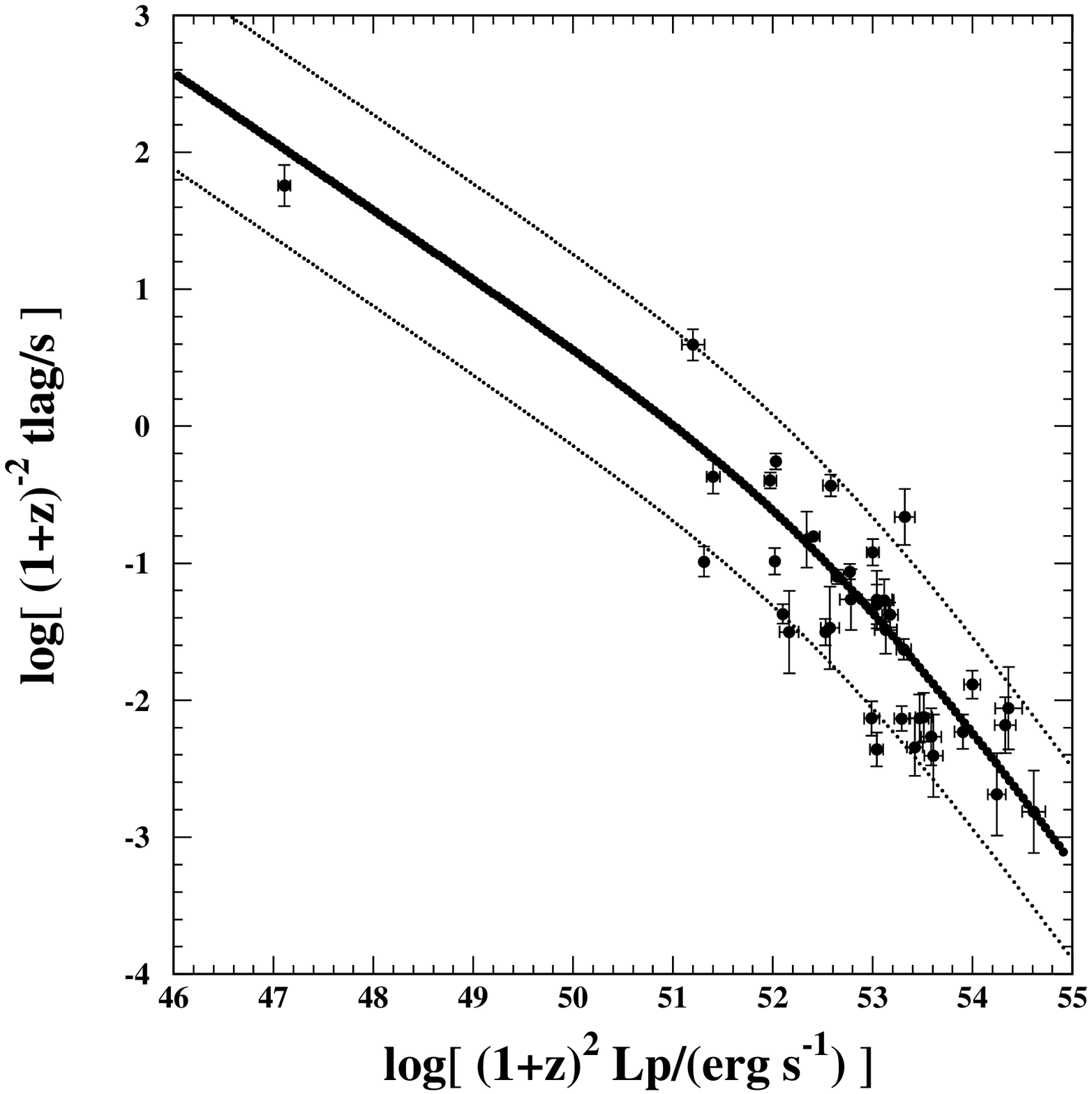,width=8cm}
 \epsfig{file=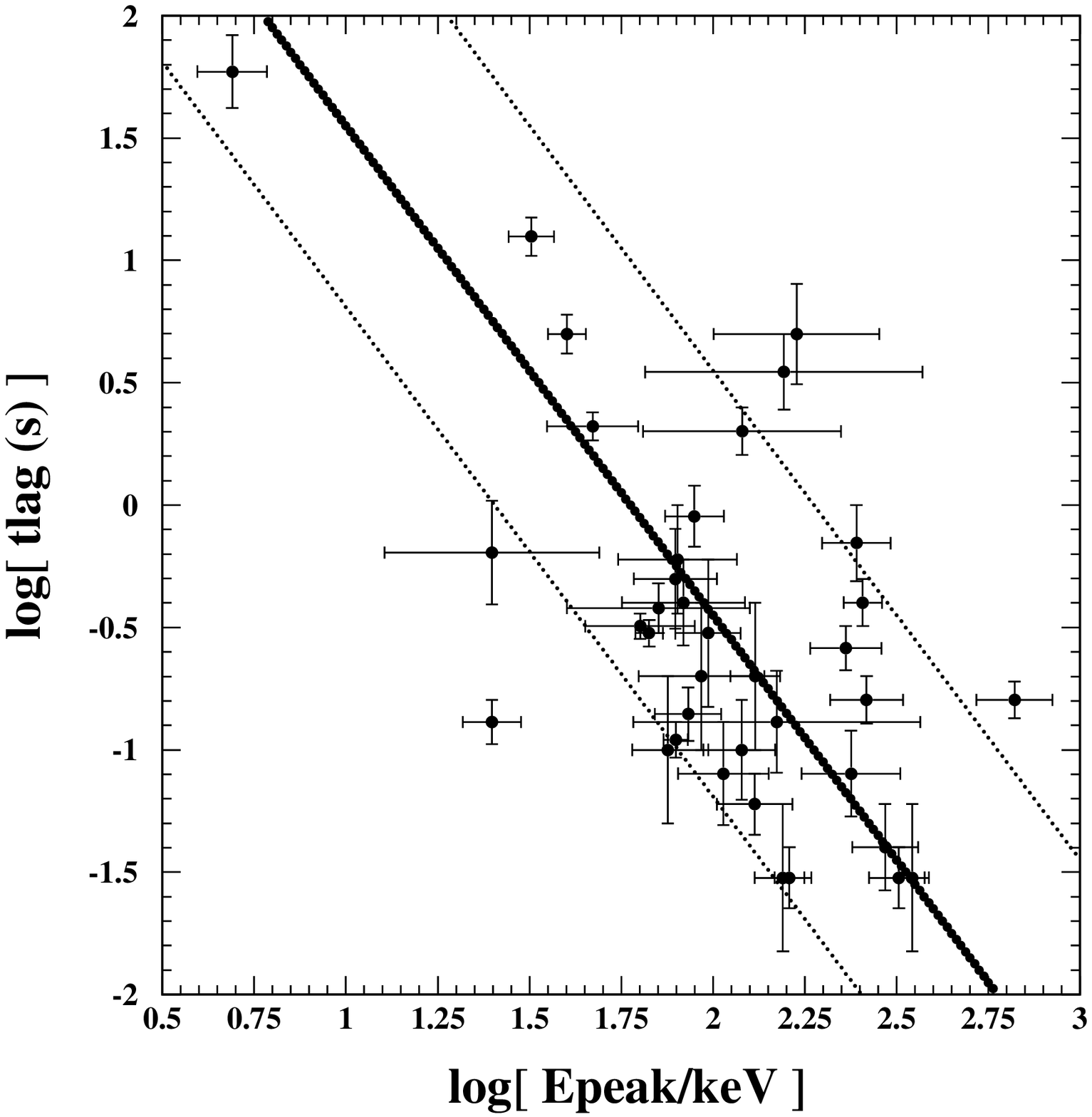,width=8cm }
}}
\vspace{-3cm}
%\hskip -.8cm
\vbox{
\hbox{
\epsfig{file=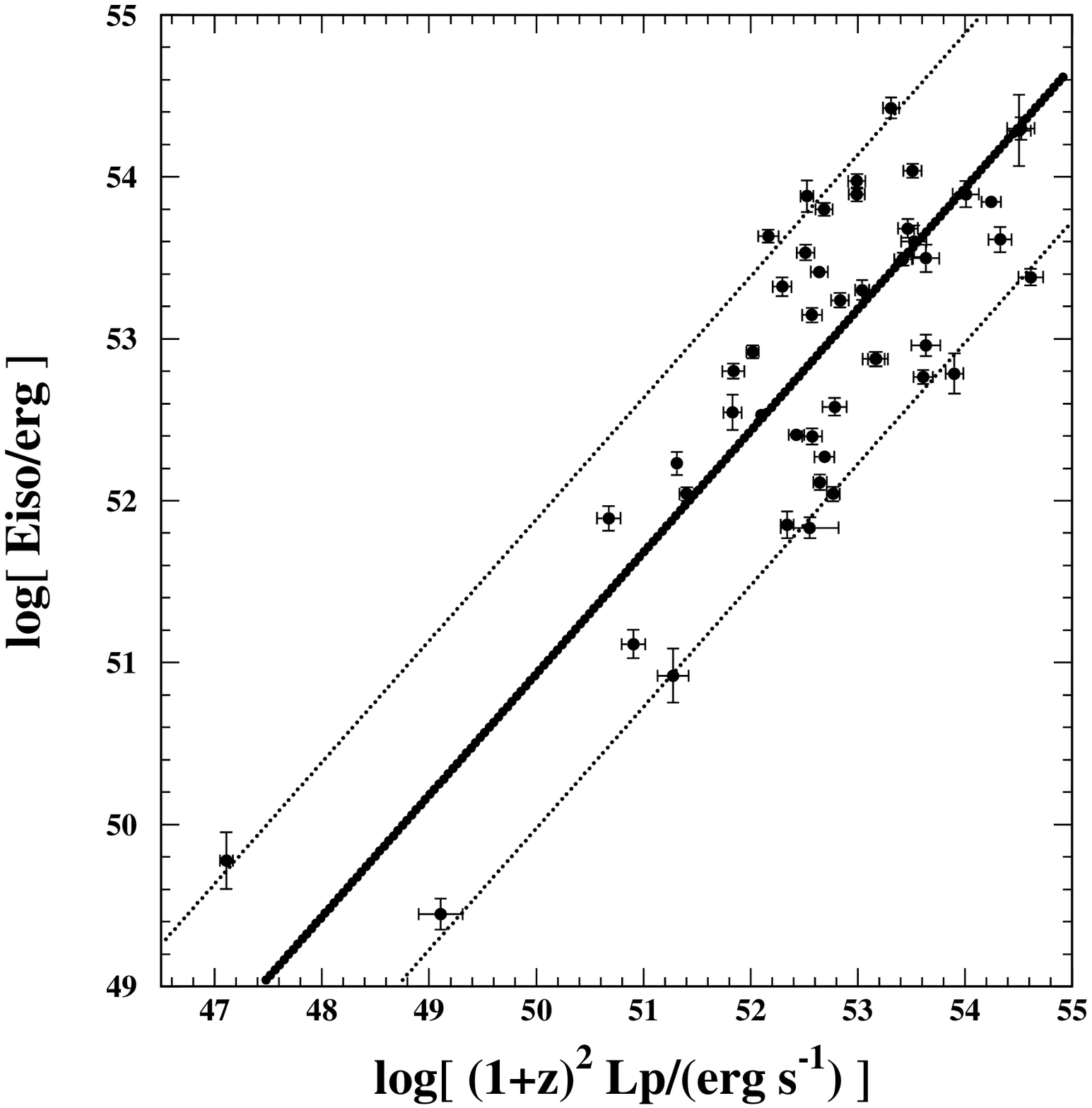,width=8.cm}
\hskip -.3cm
\vspace*{30pt}
\epsfig{file=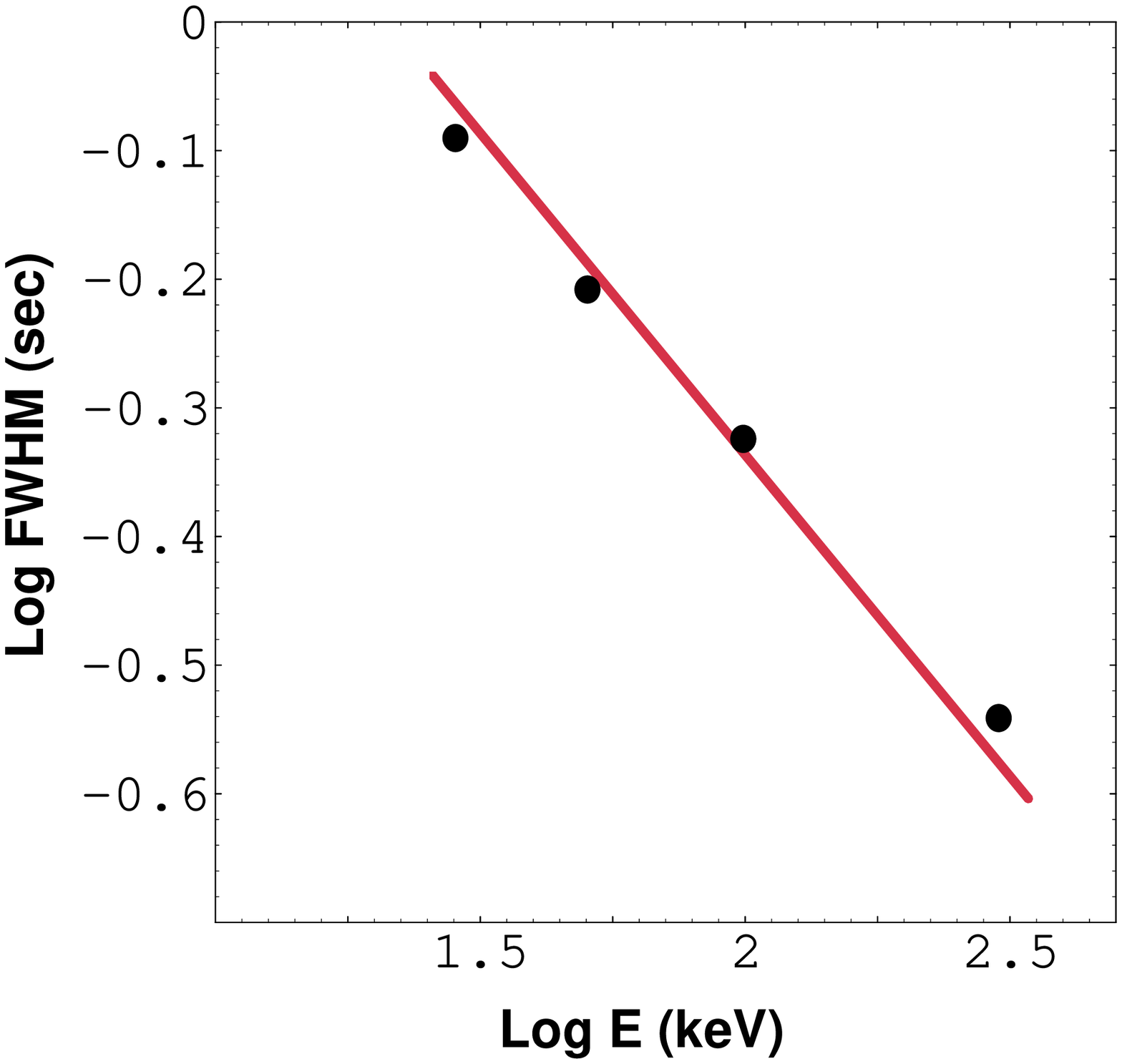,width=8.2cm}
\vspace{-1cm}
}}
\vspace*{8pt}
\caption{ Left to right and top to bottom (a to d):
a) The $[t_{\rm lag},L_p^{\rm iso}]$ correlation of Eq.~(\ref{tlageq}).
b) The $[t_{\rm lag},E_p]$ correlation of Eq.~(\ref{simple}).
c) The $[E_\gamma^{\rm iso},L_p^{\rm iso}]$ correlation of
Eq.~(\ref{simple}).
 d) The correlation between $t_{\rm FHHM}$ and BATSE 
$E$-channel of Section \ref{fwhm}.
The dotted `variance lines' are to guide the eye, they are not
always symmetric about the best fit.
}
 \label{f2}
\end{figure}

The  predictions of Eqs.~(\ref{epi}) to (\ref{tlageq}) are 
tested in Figs.~\ref{f1} and \ref{f2}a. The data are 
from Schaefer (2006); values of $E_0^{\rm iso}$
and data for some extra GRBs at low $E_0^{\rm iso}$ and $E_p$
(which could be classified as X-ray flashes) are from Amati
(2006a,b,c). All results are satisfactory.

Two observables ($L_p^{\rm iso}$ and $t_{\rm min}^{\rm rise}$)
reflect fixed values of $\gamma$ and $\delta$, since they
refer to a particular pulse in a GRB light curve, not necessarily
the same one.
For multi-pulse GRBs, the other observables reflect averages over the 
various pulses. Single-peak correlations should
be tighter than the ones we have discussed, 
but properly analized data are not available.
We also expect
correlations between two multi-pulse-averaged observables to be 
 tighter than those between a multi-pulse  observable and a 
single-pulse one.
The only correlation of the former type in Figs.~\ref{f1} and \ref{f2} is
that between $E_p$ and $E_\gamma^{\rm iso}$ in Fig.~\ref{f1}a;
it is indeed the tightest.

The correlations we have discussed are not the simplest ones the
CB model suggests. Indeed, it follows from Eqs.~(\ref{brief},\ref{tlagV})
that:
\begin{equation}
t_{\rm lag} \propto E_p^{-2},\;\;\;\;\;
E_\gamma^{\rm iso}\propto
[(1+z)^2\,L_p^{\rm iso}]^{3/4},\;\;\;\;\;
V\propto E_p,\;\;\;\;\;
t_{\rm rise}  \propto E_p^{-1}.
\label{simple}
\end{equation}
These relations involve just one
parameter: the proportionality factor, and deserve to be studied,
even if they add no significance to the results, for they are redundant 
with the correlations we have already discussed. The first two
predictions in Eqs.~(\ref{simple}) are shown in Figs.~\ref{f2}b,c.
The correlations of $V$ and $t^{\rm rise}_{\rm min}$
with $E_p$  are less informative, 
not only because the first two of these observables are of
a somewhat debatable significance, but because
the dynamical ranges of $V$, $t^{\rm rise}_{\rm min}$
and $E_p$ span $\sim 2$ orders of
magnitude, while the data on $L_p^{\rm iso}$, 
$E_\gamma^{\rm iso}$ and $t_{\rm lag}$ span $\sim 3$ times
as many.

\section{Variability and minimum rise-time}
\label{variability}

In the CB model there are two a priori time scales determining
the rise-time and duration of a pulse: the time it takes
a CB to expand to the point at which it becomes transparent
to radiation and the time it takes it to travel to a distance
from which the remaining of its path is transparent
to $\gamma$ rays (Dar \& De R\'ujula 2004). These two times
are, for typical parameters, of the same order of magnitude.
We discuss the second time scale here, for it is the one
naturally leading to
larger variabilities and differences in rise-time.   

The $\gamma$ rays of a GRB's pulse must traverse the
pre-SN wind material remaining upstream of their production
point, at a typical distance of $r={\cal{O}}(10^{16}\,\rm cm)$ 
from the parent SN. At these `short' distances, the observed  
circumburst material is located in layers whose density
decreases roughly
as $1/r^2$ and whose typical $\rho\,r^2$ is large: 
$\sim 10^{16}\,{\rm g \,cm}^{-1}$
(Chugai et al.~2003; Chugai \& Danzinger 2003).
Compton absorption in such a wind implies that a pulse of
a GRB initially rises with time as ${\rm Exp}[-(t^w_{tr}/t)^2]$, where
(Dar \& De R\'ujula 2004):
\begin{equation}
{t^w_{tr}\over 1+z}=(0.13\,{\rm s})\;
{\rho\,r^2\over 10^{16}\,{\rm g \,cm}^{-1}}\;
 {10^6\over \gamma\,\delta}.
\label{twtrans22}
\end{equation}
 The values of $\delta\,\gamma$
may differ for the different CBs (pulses) of a GRB
even if they are emitted in precisely the same direction, which 
need not be the case, e.g. if the emission axis precesses. 
The minimum rise-time, $t_{\rm rise}$,
used as a variability measure in Schaefer (2006), satisfies
Eq.~(\ref{twtrans22}), and was used in Fig.~\ref{f1}c.

The result in Eq.~(\ref{twtrans22}) is for an ideal spherically-symmetric 
wind. Actual wind distributions are layered and patchy,
implying an in-homogeneous distribution of the glory's light density.
Since the number of photons Compton up-scattered by the CB 
is proportional to this density, the
inhomogeneities would directly translate into a variability on top
of a smooth pulse shape, which reflects the average density distribution
of the wind-fed medium. 
This corresponds to a source of variability
that, as a function of $\delta$, $\gamma$ and $z$,
 behaves as the inverse of $t^w_{tr}$, the form used for $V$
in Eq.~(\ref{brief}) and Fig.~\ref{f1}d 
(in the CB model, the deviations from a smooth
behaviour observed in some optical and X-ray AGs  also trace
the inhomogeneities in the density of the interstellar 
medium; see Dado et al.~2003a, 2006). There are many ways to define the
variability of a GRB;  variations
of the sort we have described are the ones
 studied by Schaefer (2006). The data in
Figs.~\ref{f1}c,d are from his analysis and definitions.

\section{The lag-time}
\label{lag}

In the CB model  a pulse's $\gamma$-ray number flux
as a function of energy and time is of the form (Dar \& De R\'ujula, 2004):
\begin{equation}
N(E,t)\equiv{d^2N\over dE\,dt}={dN_1(E,t)\over dE}\,{dN_2(t)\over dt}\, .
\label{NEt}
\end{equation}
The function $dN_2/dt$
is well approximated by 
${\rm Exp}[-(t^w_{tr}/t)^2]\,\{[1-{\rm Exp}[-(t^w_{tr}/t)^2]\}$,
with $t^w_{tr}$ as in Eq.~(\ref{twtrans22}).
The predicted shape of $dN_1/dE$ is amazingly similar to that of 
`Band's' phenomenological spectrum and has a weak time-dependence that makes the 
spectrum within a pulse soften with time, in a time  of
order $t^w_{tr}$. The energies $E$ in $dN_1/dE$ scale with $T$:
\begin{equation}
T\equiv {4\over 3} \;T_i\;{\gamma\;\delta\over 1+z}\;\,
\langle 1+\cos\theta_i\rangle,
\label{Teff}
\end{equation}
where $\theta_i$ is the angle of incidence of a glory's photon onto
the CB (in the SN rest system) and $T_i$ is the pseudo-temperature
in the thin thermal-bresstrahlung spectrum $[{\rm Exp}(-E_i/T_i)]/E_i$
of the glory's light. We conclude that
$N(E,t)=F\left({E/T}\; , {t/ t^w_{tr}}\right)$
with $F$ a predicted function. This implies that
\begin{equation}
t_{\rm rise}(E)=t^w_{tr}\; G(E/T),
\label{tpformula}
\end{equation}
with $G(x)$ a slowly-varying function of $x$
(a fact that can be traced back to the time dependence
of $dN_1/dE$ being much slower than that of $dN_2/dt$).

Let $t_{\rm rise}(E_i)$ be the rise-time of a pulse  at a given $\gamma$-ray energy $E_i$. The lagtime is defined and approximated as:
\begin{equation}
t_{\rm lag}\simeq t_{\rm rise}(E_2)-t_{\rm rise}(E_1) \approx \Delta\,E\;{dt_{\rm rise}\over dE}\; ,
\label{tlagdef}
\end{equation}
where $\Delta E$ is generally taken to be a fixed
energy interval between two `channels'
in a given detector.
Use Eqs.~(\ref{tpformula}, \ref{tlagdef}) to deduce that:
\begin{equation}
t_{\rm lag}\approx t^w_{tr}\,{dG\over dE}\,\Delta\,E\propto
 {t^w_{tr}\over T}\,\Delta\,E
\label{tlagr1}
\end{equation}
where, on dimensional grounds, we used $dG/dE\propto 1/T$.
It follows from Eqs.~(\ref{Teff},\ref{twtrans22}) that:
\begin{equation}
t_{\rm lag}\propto 
 {t^w_{tr}\over T} \propto {(1+z)^2\over \delta^2 \, \gamma^2}\,,
\label{tlagr2}
\end{equation}
the result announced in Eq.~(\ref{tlagV}) and used in Fig.~\ref{f2}a.

\section{The duration of a GRB pulse}
\label{fwhm}

Some correlations do not follow from comparisons of $\gamma$
and $\delta$ dependences. One of them (Dar \& De R\'ujula 2004)
is the following. As the CB reaches the more transparent
outskirts of the wind, its ambient light becomes
increasingly radially directed, so that the average $1+\cos\theta_i$ in 
Eq.~(\ref{Teff}) will  tend to $0$ 
as $1/r^2\propto 1/t^2$. 
 Since the (exponential) rise of a 
typical pulse is much faster than its
(power) decay, the width of a peak is dominated by its late
behaviour at $t>t_{tr}$. At such times, $T\propto 1/t^2$ in Eq.~(\ref{Teff}),
so that $dN/dE$ is, approximately, a function of the combination $E\,t^2$.
Consequently, the width of a GRB pulse in different energy bands is:
$ \Delta t\propto E^{-1/2}$,  
in agreement with the observation,
$t_{_{\rm FWHM}}\propto E^{-0.43\pm 0.10}$ ,
for the average FWHM of peaks as a function of the energies
of the four BATSE channels (Fenimore et al.~1995, Norris et al.~1996).
This correlation is shown in Fig.~\ref{f2}d.

\section{Conclusions}

The CB model is very successful in its description of all properties
of the pulses of long-duration GRBs (Dar \& De R\'ujula 2004).
We have extended our previous discussions 
of one of these properties: the correlations between pairs of observables. 
We have analyzed a wealth of newly available data and derived
predictions for some observables which we had not studied before.
Although our predictions are expected to be better satisfied for individual pulses, the results are very satisfactory
even when applied to entire GRBs: all of the 
predicted trends agree with the observations. The
correlations we have discussed have a common and simple physical basis:
relativistic kinematics and Compton scattering. The viewing
angle $\theta$  is the most crucial parameter underlying the
correlations, and determining the properties of GRBs and 
their larger-$\theta$ counterparts, X-ray flashes
(Dar \& De R\'ujula 2004, Dado et al.~2004).

\noindent
{\bf Acknowledgements.}
{  This research was supported in part by the
Helen Asher Space Research Fund and the Institute for
Theoretical Physics at the Technion Institute. 
ADR thanks the Institute for its hospitality.}

\section{Appendix}

In this Appendix, and in the example of the $[E_p,\,E_\gamma^{\rm iso}]$
correlation, we prove in detail the ``double-power" nature of many of
the correlations predicted by the CB model.

For a typical angle of incidence (Dar \& De R\'ujula 2004), the
energy of a Compton up-scattered photon from the SN glory
is Lorentz and Doppler boosted by a factor $\sim\!\gamma\,\delta/2$
and redshifted by $1\!+\! z$. The peak energy $E_p$ of the GRB's 
$\gamma$-rays
is related to the peak energy, $\epsilon_p\!\sim\! 1$ eV, of the
glory's light by:
\begin{equation}
 (1+z)\,E_p\simeq {\gamma\,\delta\, \epsilon_p\over 2}\simeq
 (500\;{\rm keV})\; {\gamma\,\delta\over 10^6}\,
{\epsilon_p\over 1\;\rm eV}\; .
 \label{eobs}
\end{equation}  
%The upscattered radiation,  emitted nearly isotropically 
%in the CB's rest frame, is boosted by its highly relativistic motion
%to a narrow  angular distribution whose number density is:
%\begin{equation}
%{dn_\gamma \over d\Omega}\simeq {n_\gamma \over 4\, \pi}\, \delta^2 
%                       \simeq {n_\gamma \over 4\, \pi}\, {4\, \gamma^2
%                       \over (1+\gamma^2\, \theta^2)^2}\, ,
%\label{beaming} 
%\end{equation} 
The spherical equivalent
energy, $E_\gamma^{\rm iso}$, is  (Dar \& De R\'ujula 2004, Dado et al.~2006b): 
\begin{equation} 
E_\gamma^{\rm iso} \simeq 
{\delta^3\, L_{_{\rm SN}}\,N_{_{\rm CB}}\,\beta_s\over 6\, c}\,
                      \sqrt{\sigma_{_{\rm T}}\, N_b\over 4\, \pi}\sim
                      (3.8\! \times\! 10^{53}\,{\rm erg})\,{\delta^3\over 10^9}\,
{L_{_{\rm SN}}\over L_{_{\rm SN}}^{\rm bw}}\,{N_{_{\rm CB}}\over 6}\,
\beta_s\sqrt{ N_b\over 10^{50}}\; ,
\label{eiso} 
\end{equation} 
where $L_{_{\rm SN}}$ is the mean SN optical luminosity just 
prior to the ejection of  CBs, $N_{_{\rm CB}}$ is the number of CBs in 
the jet, $N_b$ is their mean baryon number, $\beta_s$ is the comoving early
expansion velocity of a CB (in units of $c/\sqrt{3}$), 
and $\sigma_{_{\rm T}}$ is the Thomson cross section. The early SN luminosity required to 
produce the mean isotropic energy, $E_\gamma^{\rm iso}\!\sim\! 4\!\times\! 10^{53}$ 
erg, of ordinary long GRBs is 
$L_{_{\rm SN}}^{\rm bw}\!\simeq\! 5\!\times\! 10^{42}\, {\rm erg\, 
s^{-1}}$, the estimated early luminosity of SN1998bw. 

\begin{figure}[]
\centering
\vbox{
\hbox{
%\vspace{2cm}
 \epsfig{file=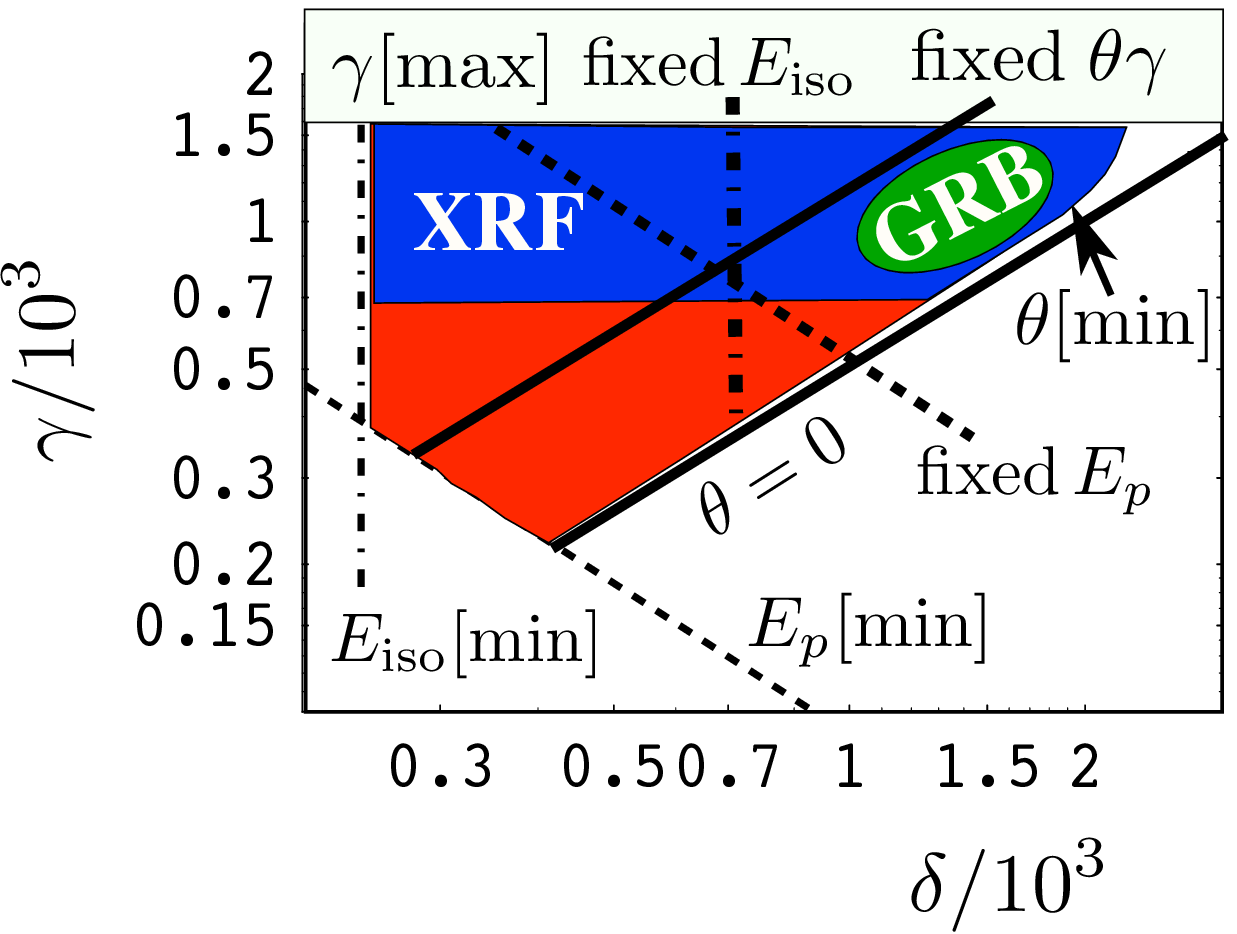,width=8cm}
 \hspace{1.cm}
 \epsfig{file=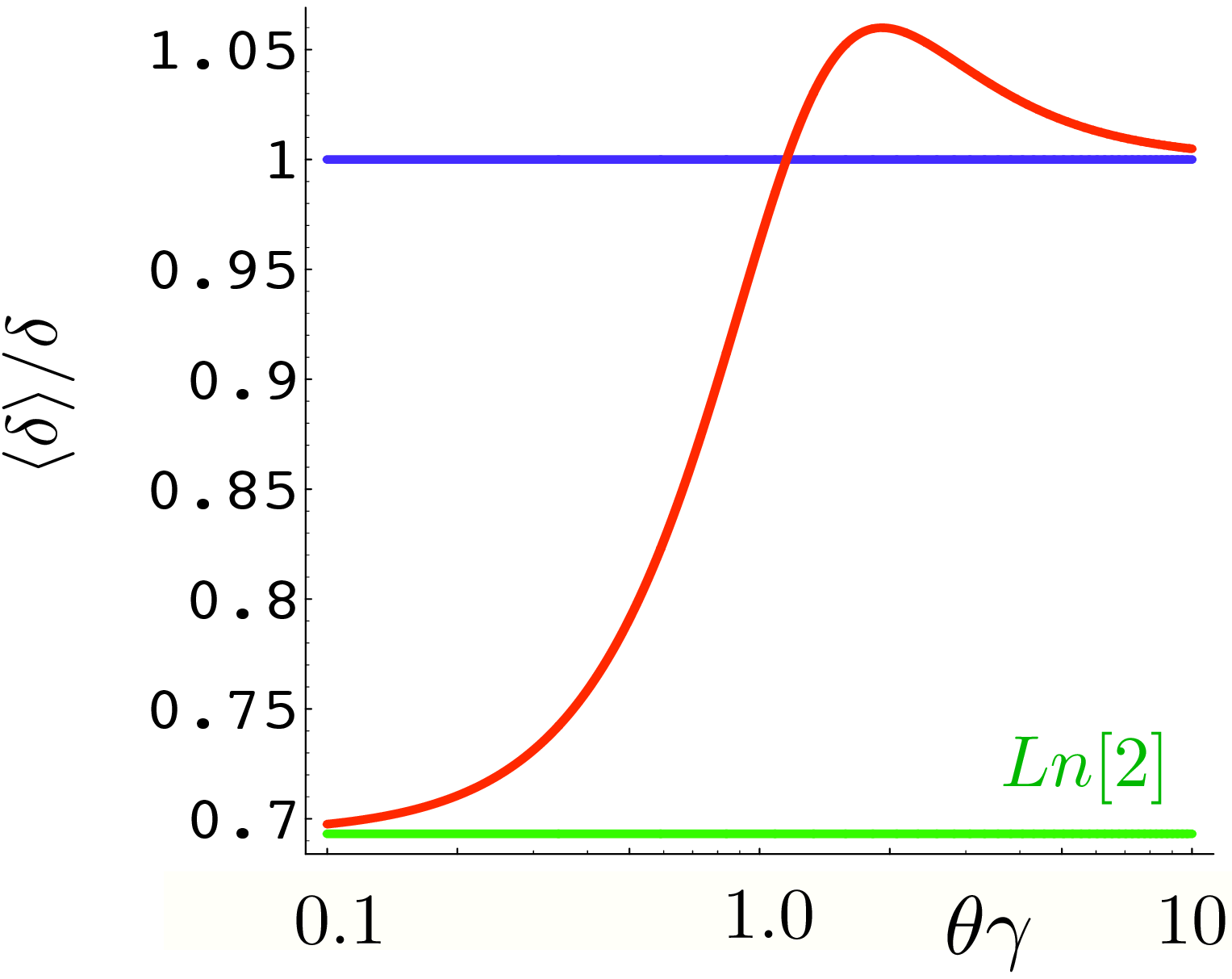,width=7.5cm}
}
\vspace{2cm}
\hbox{
\epsfig{file=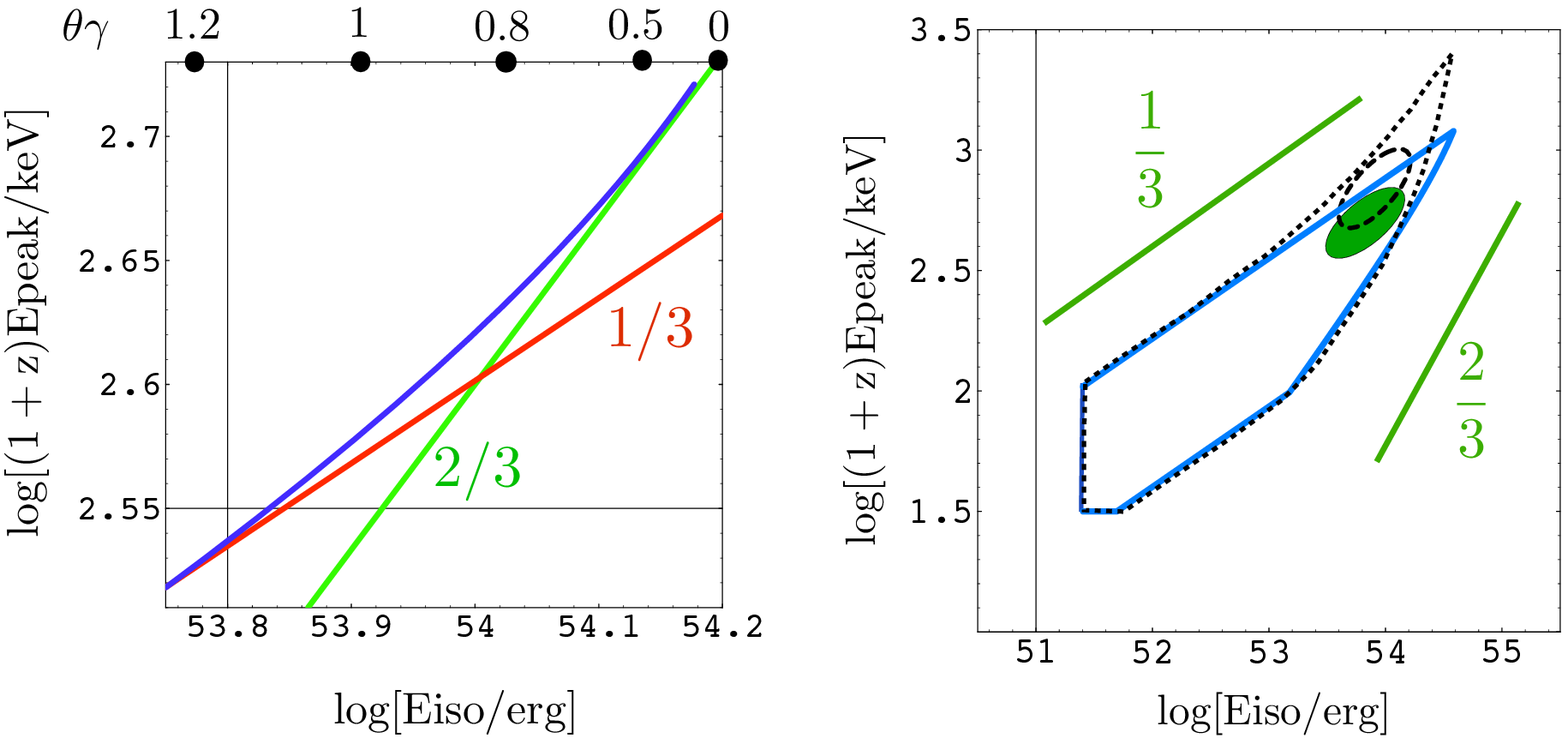,width=16cm}
%\vspace*{-50pt}
%\epsfig{file=EpEisoCurved2.eps,width=7.5cm}
}}
\vspace*{8pt}
\caption{
Left to right and top to bottom: a) The $[\delta,\,\gamma]$ domain.
b) The ratio $\langle\delta\rangle/\delta$ of average to naive Doppler
shifts for a (Lorentz-contracted)
disk-like CB of angular size $1/\gamma$ (in the SN rest
system), as a function of $\theta\,\gamma$. c) $E_p$ versus 
$E_\gamma^{\rm iso}$ for the same CB, at fixed $\gamma$.
d) Contours of $E_p$ versus $E_\gamma^{\rm iso}$ for an ensemble of CBs
whose values of $[\delta,\,\gamma]$ are in the blue domain of
Fig.~\ref{f3}a. The continuous (blue) contour does not take into account
the non-point-like character of CBs. The dashed contour does.
}
 \label{f3}
\end{figure}

The explicit proportionality factors in the relations
  $E_p\propto\gamma\,\delta$
 and $E_\gamma^{\rm iso}\propto\delta^3$ are given by
 Eqs.~(\ref{eobs},\ref{eiso}). Let us first consider them fixed
 at their typical values.
 The typical $[\gamma,\,\delta]$ domain of observable GRBs is 
 then the one shown in 
 Fig.~\ref{f3}a. The observed values of $\gamma$ are fairly
 narrowly distributed around $\gamma\!\sim\!10^3$ (Dado et al.~2003a,
 Dar \& De R\'ujula, 2004),
 as in the blue strip of the figure. The $[\gamma,\,\delta]$ domain
 is also limited by a minimum
 observable isotropic energy or fluence (both $\propto\,\delta^3$), by
 a minimum observable peak energy, and by the line
$\theta=0$ or, if one takes into account that phase space for
 observability diminishes as $\theta\to 0$, by a line corresponding to
 a minimum fixed $\theta$. The elliptical ``sweet spot" in  Fig.~\ref{f3}a
 is the region wherein GRBs are most easily detectable, particularly
  in pre-Swift times. In the CM model X-ray Flashes are GRBs
 seen at a relatively large $\theta\gamma$ (Dar \& De R\'ujula, 2004, 
 Dado et al.~2004) they populate the region labeled
 XRF in the figure, above the fixed $\gamma\theta$ line or to the
 left of the fixed $E_p$ line.

 The blue line in  Fig.~\ref{f3}d is the contour of the blue domain of 
 Fig.~\ref{f3}a,
 shown in the $[\gamma\delta,\,\delta^3]$ plane of the 
 $[E_p,\,E_\gamma^{\rm iso}]$ correlation. At low values of these quantities,
 the correlation is $E_p\,\propto\!(E_\gamma^{\rm iso})^{1/3}$. At the opposite
 extreme, the expected power is half-way
 between 1/3 and 2/3. But there is another effect
 increasing this expectation to $\sim\!2/3$.
 
A CB that is expanding, in its rest system, at a speed of relativistic sound
($\beta_{\rm exp}\!=\!\beta_s$)
--or at the speed of light ($\beta_{\rm exp}\!=\!1$)-- 
subtends a non-vanishing angle from its
point of emission. In the SN rest system, this (half-)angle is 
$\theta_{_{\rm CB}}\!=\!\beta_{\rm exp}/\gamma$. At a fixed observer's angle,
$\theta$,
the value of $\delta$ and $\delta^3$ entering Eqs.~(\ref{eobs},\ref{eiso})
are not the ``naive" ones of Eq.~(\ref{delta}), but are averages,
$\langle\delta\rangle$ and $\langle\delta^3\rangle$, over the
CB's non-vanishing surface. In Fig.~\ref{f3}b we show function 
$\langle\delta(\theta\gamma)\rangle$, for fixed $\gamma$ and 
$\theta_{_{\rm CB}}\!=\!1/\gamma$ (the result is very similar for 
$\beta_{\rm exp}=1/\sqrt{3}$). This figure is easy to interpret: for
$\theta\gg\theta_{_{\rm CB}}$, the CB is effectively point-like and
$\langle\delta(\theta\gamma)\rangle\to \delta$. At the opposite
extreme, $\theta\ll\theta_{_{\rm CB}}$, the observer's angle is immaterial
and $\langle\delta(\theta\gamma)\propto\gamma$. The consequences
of this fact on the $[E_p,\,E_\gamma^{\rm iso}]$ correlation
can be seen in Fig.~\ref{f3}c, where we have plotted 
$E_p\!\propto\!\gamma\langle\delta\rangle$ versus 
$E_\gamma^{\rm iso}\!\propto\!\langle\delta^3\rangle$, at fixed $\gamma$.
The $E_p(E_\gamma^{\rm iso})$ functional
dependence smoothly evolves from a 1/3 to a 2/3 power. 
We plot in Fig.~\ref{f3}d, as the banana-like
dashed line, the border of the blue 
domain of Fig.~\ref{f3}a, taking into account the geometrical effect we just
described.  The result is an $[E_p,\,E_\gamma^{\rm iso}]$ correlation
with an index varying from 1/3 to $\sim\!2/3$.

Finally, we may consider the effect of varying the proportionality factors
in Eqs.~(\ref{eobs},\ref{eiso}) around their reference values. This results
in a superposition of banana-like domains, the general behaviour of
which we have approximated by the $[E_p,\,E_\gamma^{\rm iso}]$ correlation
of Eq.~(\ref{epi}). Similar considerations apply to all the other two-power
correlations that we have studied.



\begin{thebibliography}



\bibitem[2006]{AM2006a}
Amati, L., 2006a, MMRAS,  372, 233
%\bibitem[2006]{AM2006b}
%Amati, L.,??? Pian 2006b, astro-ph/0607148 
\bibitem[2006]{AM2006b}
Amati, L. 2006b, astro-ph/0601553
\bibitem[2006]{AM2006c}
Amati, L., 2006c, astro-ph/0611189

\bibitem[2003]{ChugaiD}
Chugai, N.N. \&  Danziger, I.J.~2003, Astron. Lett. 29  649

\bibitem[2003]{Chugaietal}
Chugai, N.N. et al.
Proceedings of IAU Colloquium 192, {\ it Supernovae}, 
eds. J. M. Marcaide and K. W. Weiler

\bibitem[2002]{DDD2002}
Dado, S., Dar, A. \& De R\'ujula, A.~2002, A\&A, 388, 1079

\bibitem[2003]{DDD2003a}
Dado, S., Dar, A. \& De R\'ujula, A.~2003a, A\&A, 401, 243

\bibitem[2003]{DDD2003b}
Dado, S., Dar, A. \& De R\'ujula, A.~2003b, ApJ, 594, L89

\bibitem[2004]{DDD2004}
Dado, S., Dar, A. \& De R\'ujula, A.~2004, A\&A, 422, 381

%\bibitem[2004]{DDD2004b}
%Dado, S., Dar, A. \& De R\'ujula, A.~2004b, 

%\bibitem[2006]{DDD2006}
%Dado, S., Dar, A. \& De R\'ujula, A.~2006, ApJ. 646, L2106325

%\bibitem[2006]{DDD2006}
%Dado, S., Dar, A. \& De R\'ujula, A.~2006, ApJ. 646, L21

\bibitem[2003]{DDD2006a}
Dado, S., Dar, A. \& De R\'ujula, A.~2006a, ApJ, 646, L21

\bibitem[2006b]{DDDP2006b}
Dado, S., Dar, A., De R\'ujula and Plaga, R., astro-ph/0611161

%\bibitem[2005]{DD2005}

%Dado, S. \& Dar, A. 2005 ApJ. 627  L109

%\bibitem[2005]{DD2005a}
%Dado, S. \& Dar, A., 2005a, Nuovo Cimento 120, 731

%\bibitem[2004]{DAR2004}
%Dar, A. 2004, Proc. 2004 Vulcano Workshop  (eds. F. Giovannelli
%\& G. Mannocchi) p. 287 (astro-ph/0405386) 

\bibitem[2000]{DP1999}
Dar, A. \& Plaga, R. ~1999, A\&A, 349, 259

\bibitem[2000]{DD2000}
Dar, A. \& De R\'ujula, A.~2000, astro-ph/0008474

\bibitem[2000]{DD2001}
Dar, A. \& De R\'ujula, A.~2001, astro-ph/0012227

\bibitem[2004]{DD2004}
Dar, A. \& De R\'ujula, A.~2004, Physics Reports, 405, 203

%\bibitem[2004]{DD2006}
%Dar, A. \& De R\'ujula, A.~2006, hep-ph/0606199

%\bibitem[1992]{DKNR1992}
%Dar, A., Kozlovsky, B.; Nussinov, S. \& Ramaty, R. 1992, ApJ, 388, 164

\bibitem[1987]{DR1987}
De R\'ujula, A, 1987, Phys. Lett. 193, 514

\bibitem{}
Fenimore, E. E., in 't Zand, J. J. M., Norris, J. P., Bonnell, J. T.
\& Nemiroff, R. J. 1995, ApJ, 448, L101

\bibitem[2000]{FenRR2000}
Fenimore, E.E. \& Ramirez-Ruiz, E. 2000, astro-ph/0004176

\bibitem[2006]{Guid}
Guidorzi, C., et al.  2006, MNRAS, 371, 843


\bibitem{}
Norris, J. P., Nemiroff, R. J., Bonnell, J. T., Scargle, J. D.,
Kouveliotou, C., Paciesas, W. S., Meegan, C. A. \& Fishman, G. J. 1996,
ApJ, 459, 393

\bibitem[2001]{Plaga}
Plaga, R. 2001, A\&A, 370, 351

\bibitem[2001]{Reichart}
Reichart, D. E., et al.  2001, ApJ, 552, 57

\bibitem[2006]{WB2006}
Schaefer, B. E., astro-ph/0612285  

\bibitem[1995]{SD1995}
Shaviv, N. J., Dar, A. 1995, ApJ, 447, 863


\bibitem[2004]{YO2004}
Yonetoku, D., et al. 2004, ApJ. 609, 935


\end{thebibliography}
\end{document}